# On scattering cross sections and durations near an isolated compound-resonance, distorted by the non-resonant background, in the center-of-mass and laboratory systems.

V.S.Olkhovsky, M.E.Dolinska and S.A.Omelchenko

Institute for Nuclear Research of NASU, Kiev-0650, Ukraine, olkhovsky@mai..ru

Abstract: During last 20 years there was revealed and published the phenomenon of the appearing of the time advance instead of the time delay at the region of a compound-nucleus resonance, distorted by the non-resonant background (in the center-of-mass (*C*-) system). This phenomenon is usually accompanied by a minimum in the cross section near the same energy. Here we analyze the cross section and the time delay of the nucleon-nucleus scattering in the laboratory (*L*-) system. In the *L*-system the delay-advance phenomenon does not appear. We use and concretize the non-standard analytical transformations of the cross section from the *C*-system to the *L*-system, obtained in our previous papers. They are illustrated by the calculations of energy dependences of cross sections in the *L*-system for several cases of nucleon elastic scattering by nuclei $^{12}C$, $^{16}O$, $^{28}Si$, $^{52}Cr$, $^{56}Fe$ and $^{64}Ni$ at the range of distorted resonances in comparison with the experimental data.

## 1. Introduction and the pre-history of the problem.

*The appearing of the delay-advance phenomenon in the **C**-system.* There was revealed and published in [1-7] (for various cases) a phenomenon of the appearing of the time advance instead of the time delay at the region of a compound-nucleus resonance, distorted by the non-resonant background (in the *C*-system). This phenomenon is usually accompanied by a minimum in the cross section near the same energy. Here we analyze if such phenomenon is appearing in the *L*-system? It is found in [8,9] that the standard formulas of passing from the *L*-system to the *C*-system are not valid in the presence of two mechanisms of collisions – a prompt (direct or potential) process, when the center-of-mass is practically not shifted during the collision, and a delayed process, when the long-living decaying compound nucleus is moving in the *L*-system: There are considered the motion of the long-living decaying compound nucleus (which coincides with the motion of the center-of-mass) in the *L*-system and simultaneously practical absence of the *C*-system motion during the prompt direct process, and this circumstance is not reflected automatically in the standard pure kinematic transformations from the *L*-system into the *C*-system.

Now we shall briefly describe the appearing of the *delay-advance* in nucleon elastic scattering by nuclei near a resonance, distorted by the non-resonant background (in the *C*-system). Usually (see, for instance,[1-3]) for the channel of elastic scattering of nucleons by spherical nuclei near an isolated resonance the nucleon-nucleus scattering amplitude $F^C(E, \theta)$ in the *C*-system can be written, neglecting the spin-orbital interaction, as

$$F^C(E, \mathbf{q}) = f(E, \mathbf{q}) + f_{l, res}(E, \mathbf{q}) \qquad (1)$$

where

$$f(E,\mathbf{q}) = f_{Coul}(E,\mathbf{q}) + (2ik)^{-1}\sum_{l \ne l}(2l+1)P_l(\cos\mathbf{q})\exp(2i\mathbf{h}_l)[\exp(2i\mathbf{d}_l^b) - 1],$$

$$f_{l,res}(E,\mathbf{q}) = (2ik)^{-1}(2l+1)P_l(\cos\mathbf{q})\exp(2i\mathbf{h}_l)\left[\exp(2i\mathbf{d}_l^b)\right]\frac{E^{\bullet} - E_{res}^{\bullet} - i\Gamma/2}{E^{\bullet} - E_{res}^{\bullet} + i\Gamma/2} - 1],$$

$F_{Coul}(E,\mathbf{q})$ is the Coulomb scattering amplitude, $\mathbf{d}_l^b$ and $\mathbf{h}_l$ being the background nuclear $l$-scattering phase-shift and the Coulomb $l$-scattering phase shift, respectively, $k$ is the wave number, $\mathbf{q}$ is the scattering angle in the $C$-system (here we neglect the spin-orbital interaction and consider a rather heavy nucleus), $E^{\bullet}, E_{res}^{\bullet}$ and $\Gamma$ are the excitation energy, the resonance energy and the width of the compound nucleus, respectively. Rewriting (1) in the form

$$F^C(E, \mathbf{q}) = [A(E^* - E^*_{res}) + iB\mathbf{G}/2](E^* - E^*_{res} + i\mathbf{G}/2)^{-1} \qquad (1a)$$

where

$$A = f(E, \mathbf{q}) + (k)^{-1}(2l+1)P_l(\cos\mathbf{q})\exp(2i\mathbf{h}_l + i\mathbf{d}_l^b)\sin\mathbf{d}_l^b,$$

$$B = f(E, \mathbf{q}) + (ik)^{-1}(2l+1)P_l(\cos\mathbf{q})\exp(2i\mathbf{h}_l + i\mathbf{d}_l^b)\cos\mathbf{d}_l^b,$$

we obtain (as it was made in [1,2] on the base of ref.[9]) the following expression for the total scattering *duration* $\mathbf{t}^C(E,\mathbf{q})$

$$\mathbf{t}^C(E,\mathbf{q}) = 2R/v + h\P \arg F/\P E \circ 2R/v + \mathbf{Dt}^C(E,\mathbf{q}) \qquad (2)$$

(for the quasi-monochromatic particles which have very small energy spreads $\mathbf{DE} << E, \mathbf{G}$, when one can use *the method of stationary phase* for approaching the group velocity of the wave packet [10]); in (2) $v = hk/\mathbf{m}$ is the projectile velocity and $R$ is the interaction radius, and $\mathbf{Dt}^C$ is

$$\mathbf{Dt}^C(E,\mathbf{q}) = -(h\text{Re}\mathbf{a}/2)[(E^* - E^*_{res} - \text{Im}\mathbf{a}/2)^2 + (\text{Re}\mathbf{a})^2/4]^{-1} + \mathbf{Dt}_{res}, \qquad (3)$$

with $\mathbf{Dt}_{res} = (h\mathbf{G}/2)[(E^* - E^*_{res})^2 + \mathbf{G}^2/4]^{-1}$, $\mathbf{a} = \mathbf{G}B/A$. We stress here that the total scattering duration is defined and measured (see also ref.[9]), unlike to the cross section defined macroscopically, as the duration of the *microscopic* scattering by the interaction sphere of the radius $R$ around the compound nucleus.

From (3) one can see that, if $0 < \text{Re}\mathbf{a} < \mathbf{G}$, the quantity $\mathbf{Dt}(E,\mathbf{q})$ appears to be *negative* in the energy interval $\sim \text{Re}\mathbf{a}$ around the center at the energy $E^*_{res} + \text{Im}\mathbf{a}/2$. When $0 < \text{Re}\mathbf{a}/\mathbf{G} << 1$ the minimal delay time can obtain the value $-2\hbar/\text{Re}\mathbf{a} < 0$. Thus, when $\text{Re}\mathbf{a} \to 0^+$, the interference of the resonance and the background scattering can bring to *as much as desired large of the advance* instead of the delay! Such situation is mathematically reflected by the presence of the zero $E^*_{res} + i\mathbf{a}/2$, besides the pole $E^*_{res} - i\mathbf{G}/2$ of the amplitude $F^C(E, \theta)$, or the correspondent $T$-matrix, in the lower unphysical half-plane of the Riemann surface of the complex values of $E$. This *phenomenon* *of delay-advance* for elastic scattering was revealed only in the *C*-system in [1-3] (and in other cases in [4-7]).

*The schematic picture of a collision with two mechanisms (prompt and compound-nucleus).* Following the general approach from [12], we describe the exit channel in the collision

$$x + X \quad \circledR \quad y + Y \tag{4}$$
(incident channel)    (exit channel)

as the motion of two outcoming wave packets (see Figs.1a,b), each of which has the form like

$$c_n \exp[i(k_n r_n - e_n t/\hbar)] \, G(r_n - v_n t)$$

with $e_n = \hbar^2 k_n^2 / 2m_n$, $v_n = \hbar k_n / m_n$, $c_n$ are the normalization constants, $n=1,2$. These wave packets are practically the plane waves in the limits of the wave packets $G(r_n - v_n t)$ with the space (radial) width $\gg 1/k_n$ (for quasi-monochromatic particles), which are moving with the constant group velocities $v_n$. Their group velocities practically coincide with the kinematic velocities. At least one of them (usually $y$) moves along the macroscopic distance till the registration detector and so, for instance, $r_1 \cong v_1 \cdot t$.

Now we shall analyze the scheme between prompt direct and delayed compound-resonance processes of the collision (4) and to show their qualitative difference in the *L*-system. In Fig.1 a, b these two processes in the *L*-system are pictorially presented (they represent the prompt (direct) and the delayed compound-resonance mechanisms of the emitting *y* particle and *Y* nucleus, respectively.). The both mechanisms are *macroscopically* kinematically indistinguishable but they are *microscopically* different processes:

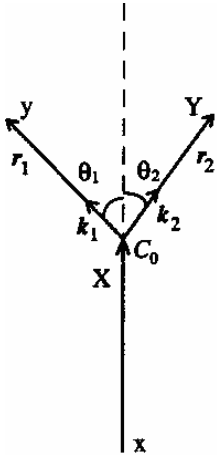  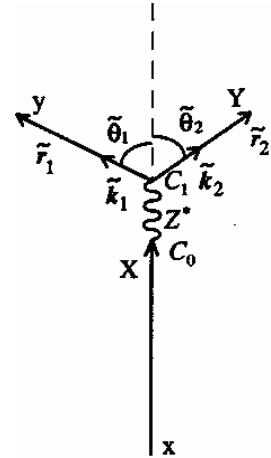

Fig.1,a                    Fig.1,b

Fig.1,a represents the direct process of the prompt emission of the final products from the collision point $C_0$ while Fig.1,b represents the motion of the compound-resonance nucleus $Z^*$ from point $C_0$ to point $C_1$, where it decays by the final products $y + Y$ after traveling a distance between $C_0$ and $C_1$ which is equal to $\sim V_C \Delta t_{res}$ before its decay. Here $V_C$ is the center-of-mass velocity and $\Delta t_{res} = (\hbar \Gamma /2) / [(E_Z - E_{res,Z})^2 + \Gamma^2/4]$ is the mean time of the nucleus $Z^*$ motion before its decay [4] when the energy spread $\Delta E$ of the incident particle $x$ is very small in comparison with the resonance width $\Gamma$ ($E_Z = E^*$, $E_{res,Z} = E^*_{res}$). For the clarity of the time difference between both processes we impose the practical evident condition

$$t_{dir} << t_{res}(E_Z) \text{ for } (E_Z - E_{res,Z})^2 \cdot 2\Gamma^2 \quad . \tag{5}$$

For the *macroscopically* defined and measured cross sections, in the case of very large macroscopic distances $r_1$ (near the detector of the final particle $y$) for very small angular and energy resolution ($\Delta q_1 << q_1$ and $\Delta k_1 << k_1$), the angles $q_1$ and $q'_1$, as well as momentums $k_1$ and $k'_1$, can be considered as practically coincident. Really, $q_1 - q'_1 \sim \Delta r_1 / r_1$ and $k_1 - k'_1 \sim \Delta r_1 / r_1$ with $|\Delta r_1| = |r_1 - r'_1|$. Using the *usual macroscopic definition of the cross section* (see, for instance, [12]) after a series of transformations for the exit asymptotic wave packet of the system $y + Y$, it was directly obtained in [8] the following expression for the cross section $\sigma$ of reaction (4) in the $L$-system in the case of quasi-monochromatic incident beam ($\Delta E << E$) and very small angular and energy resolution ($\Delta q_1 << q_1$, $\Delta E << \Gamma$) of the final-particle detector:

$$\sigma = \sigma_0^{(incoh)} + \sigma_1^{(interf)} \quad , \tag{6}$$

where

$$\sigma_0^{(incoh)} \cong \left| f_{dir}^{(L)} \right|^2 + \frac{J_{C \to L} \left| g_{Z^*}^{(C)} \right|^2}{(E_Z - E_{res,Z})^2 + \Gamma^2/4}, \tag{7}$$

$$f_{dir}^{(L)} = \sqrt{J_{C \to L}} f_{dir}^{(C)}(E_1^C, \vartheta_1^C), \; f_{dir}^{(C)} = f_C(E_1^C, \vartheta_1^C) + \tag{8}$$

$$+ \frac{1}{2ik_1^C} \sum_{l' \neq l} (2l'+1) P_{l'}(\cos \vartheta_1^C) e^{2i\eta_{l'}} (e^{2i\delta_{l'}} - 1) ,$$

$$\sigma_1^{(interf)} = 2 \left| f_{dir}^{(L)\bullet} \frac{J_{C \to L}^{1/2} g_{Z^*}^{(C)}}{E_Z - E_{res,Z} + i\Gamma/2} \right| \cos \Phi , \tag{9}$$

$$\frac{g_{Z^*}^{(L)}(E_1, E_2)}{E_Z - E_{res,Z} + i\Gamma/2} = f_{l,res}(E_1^C, \vartheta_1^C) =$$

$$\frac{1}{2ik_1^C} (2l+1) P_l(\cos \vartheta_1^C) e^{2i\eta_l} \left\{ e^{2i\delta_l} \frac{E^C - E_{res}^C - i\Gamma/2}{E^C - E_{res}^C + i\Gamma/2} - 1 \right\},$$

$$\Phi = \chi + \beta + \varphi, \; \chi = \arg(J_{C \to L}^{1/2} g_{Z^*}^{(L)}) - \arg(f_{dir}^{(L)}), \; \beta = \arg(E_Z - E_{res,Z}) + i\Gamma/2)^{-1} , \tag{10}$$

$$\phi = k_1 \boldsymbol{D} r_1 + k_2 \boldsymbol{D} r_2 \, , \quad \boldsymbol{D} r_{1,2} = V_{proj1,2} \boldsymbol{D} t_{res} \, ,$$

$V_{proj1,2}$ is the projection of the $Z^*$-nucleus velocity to the direction of $\vec{k}_{1,2}$, $f_C$ is the Coulomb scattering amplitude, $\boldsymbol{h}_l$ is the Coulomb $l$-wave partial phase shift., $\boldsymbol{d}_l$ is the $l$-wave scattering background phase shift. For the simplicity here we neglect the spin-orbital coupling and we have supposed that the absolute values of all differences $r_n/v_n - r_p/v_p$ ($n \mathbf{1} p = 1,2$) are much less than the time resolutions. Here $J_{C \circledR L}$ is the standard Jacobian of transformations from the $C$-system to the $L$-system.

### 2. The absence of the time advance near any isolated compound resonance in the $L$-system.

We underline that the obtained in [8,9] formulas (6)-(10) for the cross section $\boldsymbol{s}$, defined by the usual *macroscopic* way, takes into account a real *microscopic* motion of the compound nucleus. So, the formulas (6)-(10) evidently differ from the standard only kinematic transformation of $\boldsymbol{s}^C(E,\boldsymbol{q}) = |F^C(E,\boldsymbol{q})|^2$ from the $C$-system into the $L$-system, taking into account only the kinematic transformations of the energies and angles from the $C$-system to the $L$-system in the formal expression for $\boldsymbol{s}^C(E,\boldsymbol{q})$ without consideration of the microscopic difference between the processes in figs.1a and 1b, reflected by appearing of the parameter $\phi = k_1 \boldsymbol{D} r_1 + k_2 \boldsymbol{D} r_2$, $\boldsymbol{D} r_{1,2} = V_{proj1,2} \boldsymbol{D} t_{res.}$

Earlier (see, for instance, [1-3]) usually the analysis of the amplitudes, cross sections and durations of the elastic scattering had been made on the base of formulas (1) → (1a) in the $C$–system where the compound-nucleus motion in the $L$-system had not been taken into account. But, as it was firstly shown in [8], if one considers the motion of the decaying compound nucleus in the $L$-system, then the role of the $C$-system in the prompt (direct and potential) process and the delayed compound-nucleus process appears to be different in principle: For prompt or for the resonance compound-nucleus process the expressions for the amplitude in the $C$- and $L$-system differ not only by the standard kinematic transformations $\{E^C, \boldsymbol{q}^C\} \leftrightarrow \{E^L, \boldsymbol{q}^L\}$ but also by the motion of the decaying compound nucleus (which coincides with the motion of the $C$-system) during the time $V_C \boldsymbol{D} t_{res}$, as it is shown in figs.1a and 1b, respectively. If in [1-3] the formulas (1) and (1a) are used for the $C$-system and describe the coherent sum of interfering terms for the both cross section $\boldsymbol{s}^C(E,\boldsymbol{q}) = |F^C(E,\boldsymbol{q})|^2$ and the time delay $\boldsymbol{D}t^C(E,\boldsymbol{q})$ without taking the difference between the scattering schemes in Figs.1a and 1b (i.e. without the microscopic motion of the delaying compound nucleus from the point $C_0$ to the point $C_1$). However, *microscopically* defined (and even measured (see, for instance, [10]) time delays for the direct process (corresponding to fig.1a) and the compound-nucleus process (corresponding to fig.1b) have really negligible interference between each other (in the space-time approach): this does namely follow from the small probability of the interference term in comparison with two mean time delays of the both direct and compound-nucleus mechanisms even for the monochromatic particles, when the condition (5) is satisfied. So, really *there will no advance* in the $C$-system due to the practical absence of the *microscopic* interference between the prompt and delayed compound-nucleus processes in the case of the approximate validity of the condition (5). So, the delay-advance phenomenon in the *C*-system is really eliminated by the correct analysis of the real events in the *L*-system.

At the same time the approach (6)-(10) presents the self-consistent base for the correct analysis of the experimental data on the cross sections for the nucleon-nucleus scattering in the *L*-system. By the way, any attempt to describe the experimental data on the cross sections of the nucleon-nucleus scattering near an isolated resonance, distorted by the non-resonant background, in the *L*-system on the usual simple base of the formula (1) in the *C*-system with any variation of the scattering phases with the subsequent standard transformations $\{E^C, \mathbf{q}^C\} \leftrightarrow \{E^L, \mathbf{q}^L\}$ in the *L*-system has practically no physical sense because in this case we neglect the real compound-nucleus motion of the decaying resonance compound nucleus.

### 3. The calculations of the energy dependence of the nucleon-nucleus elastic-scattering cross section near a distorted resonance for the real description of the experimental data on the base of (6) – (10).

If in the previous paper [8] there had been presented some pure abstract examples for the simple illustrations, now we shall show the results of the calculated cross sections of the nucleon-nucleus elastic scattering, based on the comparison with the experimental data. Firstly, for the elastic proton scattering by nuclei $^{12}C$ and $^{16}O$ the calculations of the excitation function $\sigma(E)$ in the *L*-system near the distorted resonances were performed for the same values of parameters in the amplitudes of direct and resonance scattering as in [2] ($\mathbf{q}_1^L = 75.4°$, $\mathbf{d}_0 = 1.0$, $\mathbf{d}_1 = 1.2$, $\mathbf{d}_2 = 0.22$) and in [3] ($\mathbf{q}_1^L = 150°$, $\mathbf{d}_0 = \pi/8$, $\mathbf{d}_1 = -\pi/2$), respectively, and for the previously known Jacobian $J_{C \to L}$. The resonance parameters $E_{res} = 1,734$ Mev, $\Gamma = 47$ keV ($l=2$) were chosen for $p + ^{12}C$ and $E_{res} = 2,67$ MeV, $\Gamma = 14,5$ keV ($l=1$) for $p + ^{16}O$. The fitting parameter $c$ was chosen to be equal to $0.01\pi$ for $p + ^{12}C$, and to $\pi$ for $p + ^{16}O$.

The calculated by formulas (6)-(10) data in comparison with the experimental data from [2,3] for $p + ^{12}C$ and for $p + ^{16}O$ are presented in the Figures 2 and 3, respectively.

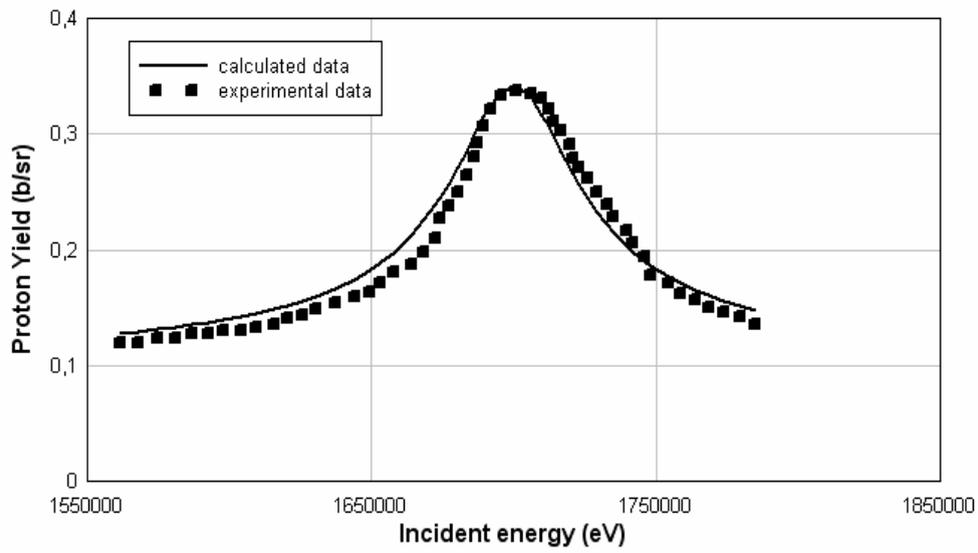

Fig. 2. Excitation function for $^{12}C$ $(p, p)$.

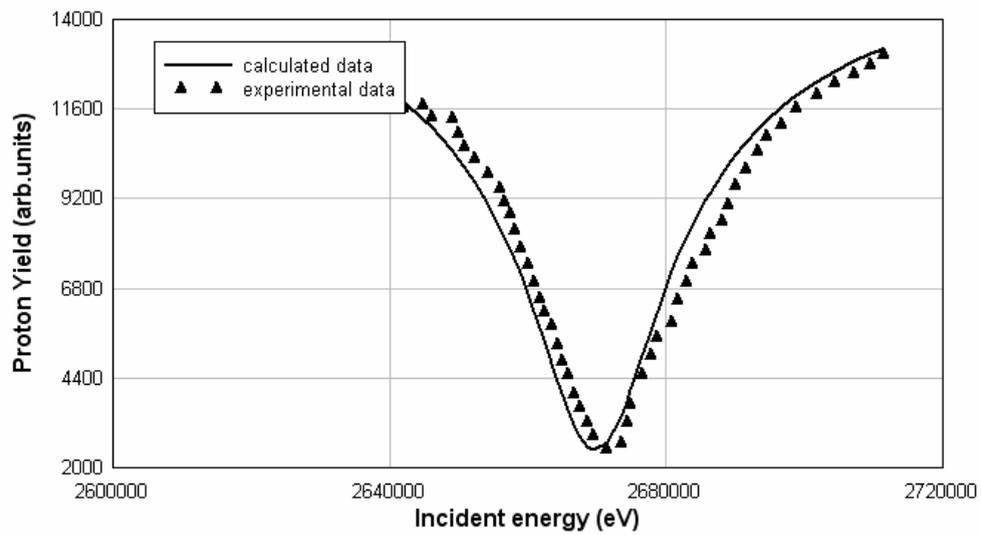

Fig. 3. Excitation function for $^{16}O$ $(p, p)$.

Then, for the very low-energy elastic neutron scattering by nuclei $^{28}Si$ or $^{52}Cr$ or $^{56}Fe$ or $^{64}Ni$ the calculations of the excitation function $s(E)$ near a distorted resonance with $E_{res}$=55.67 keV, $\Gamma$=0.48 keV or $E_{res}$=50,5444 keV, $\Gamma$= 1,81 keV or $E_{res}$=27.9179 keV, $\Gamma$=0.71 keV or $E_{res}$=24.7402 keV, $\Gamma$=0.695 keV, respectively, were performed with the simply chosen values of the parameters for the amplitudes of direct and resonance scattering in the $C$-system with $l$ =0 (and, of course, without the Coulomb phases) in the formulas (5)-(9). The fitting parameter $c$ was chosen to be equal to $0.68p$ or to $0.948\,p$ or to $0.956\,p$ or to $p$, respectively.

The calculated data in comparison with the experimental data from [13] for $n + {}^{28}Si$, $n + {}^{52}Cr$, $n + {}^{56}Fe$, $n + {}^{64}Ni$ are presented in the Figures 4-7, respectively.

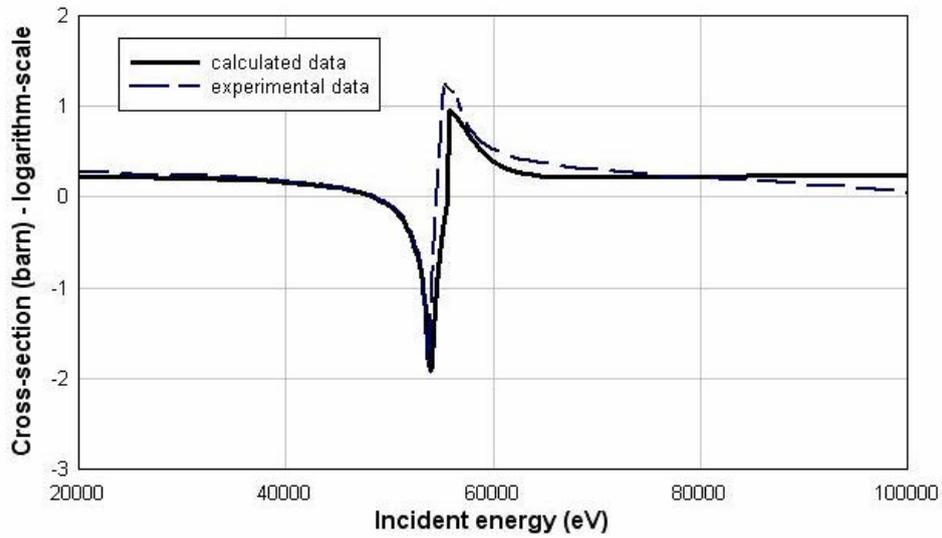

Fig.4. Excitation function for $^{28}Si$ (n, n).

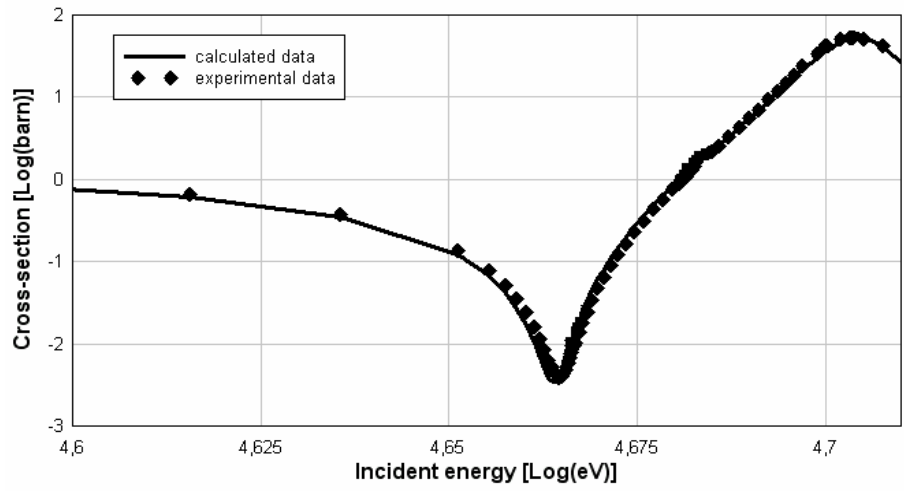

Fig.5. Excitation function for $^{52}Cr(n,n)$.

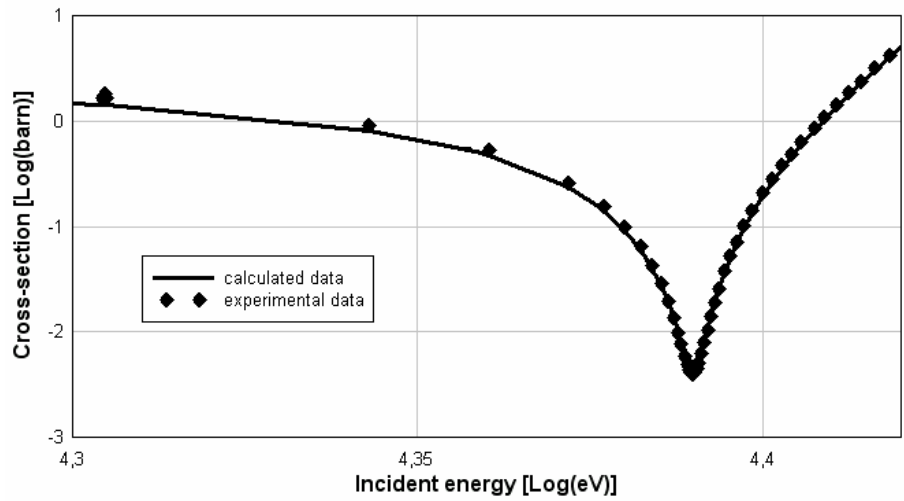

Fig.6. Excitation function for $^{56}Fe(n,n)$.

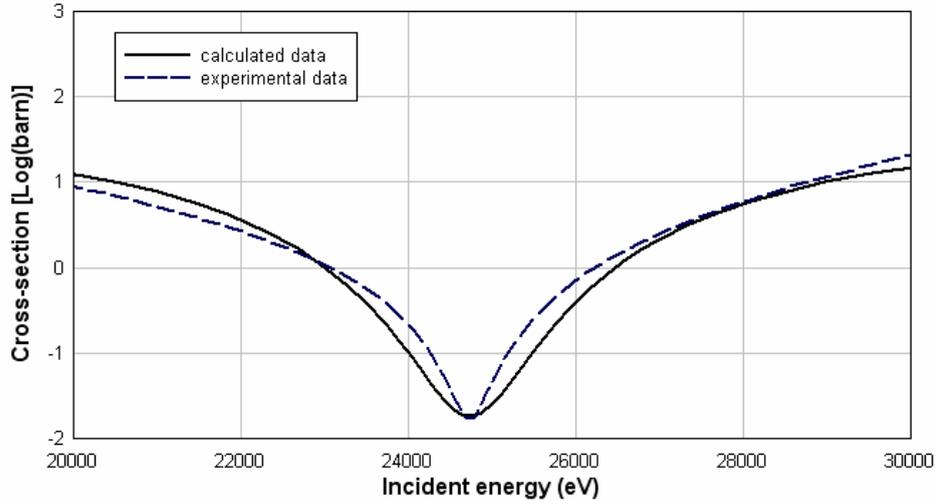

Fig.7 Excitation function for $^{64}Ni(n,n)$.

## 4. Conclusions.

(1) The *delay-advance phenomenon in the C-system*, firstly revealed for elastic scattering in [1-3], did not revealed here in the *L*-system by analysis of the time delay, using the appropriate space-time analysis of the compound-nucleus motion, following and continuing our method taken from [8,9].

(2) Unlike the pure illustrative examples in [8], being not connected with the real experimental data, here we present the new results of calculations, connected with the real experimental data for low-energy nucleon-nucleus elastic scattering in the *L*-system, taken from [2,3,13].

(3) The presented here approach can be used for a more deep extension of the analysis of all existing experimental data on the elastic nucleon-nucleus scattering in the ***L***-system, containing the resonances.

(4) The new formulas (6)-(10) can be also used for the improving of the nuclear data analysis of other two-particle channels of the nucleon-nucleus collisions and, moreover, can be generalized or changed for applications to the more complicated collisions, including those which had been presented in the above cited refs. [5-7].

**Acknowledgements**. The authors are very grateful to Dr. N.V.Eremin and to Prof. G.Giardina for the stimulation of this paper and to Dr. A.I.Kalchenko for the kind help to find various examples of neutron-nucleus cross sections.